\documentclass[apj]{emulateapj}

\usepackage{mathtools}
\usepackage{times}

\shorttitle{The Temperature Structure of the Solar Corona}
\shortauthors{Warren, Byers, \& Crump}

\slugcomment{To be submitted to the Astrophysical Journal}

\begin{document}


\title{Sparse Bayesian Inference and the Temperature Structure of the Solar Corona}

\author{Harry P. Warren\altaffilmark{1}, Jeff M. Byers\altaffilmark{2}, and
  Nicholas A. Crump\altaffilmark{3}}
\affil{\altaffilmark{1} Space Science Division, Naval Research Laboratory, Washington, DC 20375}
\affil{\altaffilmark{2} Materials Science and Technology Division, Naval Research Laboratory,
  Washington, DC 20375}
\affil{\altaffilmark{3} Naval Center for Space Technology, Naval Research Laboratory, Washington,
  DC 20375} 


\begin{abstract}
  Measuring the temperature structure of the solar atmosphere is critical to understanding how it
  is heated to high temperatures. Unfortunately, the temperature of the upper atmosphere cannot be
  observed directly, but must be inferred from spectrally resolved observations of individual
  emission lines that span a wide range of temperatures. Such observations are ``inverted'' to
  determine the distribution of plasma temperatures along the line of sight. This inversion is
  ill-posed and, in the absence of regularization, tends to produce wildly oscillatory solutions.
  We introduce the application of sparse Bayesian inference to the problem of inferring the
  temperature structure of the solar corona. Within a Bayesian framework a preference for solutions
  that utilize a minimum number of basis functions can be encoded into the prior and many ad hoc
  assumptions can be avoided.  We demonstrate the efficacy of the Bayesian approach by considering
  a test library of 40 assumed temperature distributions.
\end{abstract}



\section{Introduction}

Making detailed measurements of the temperature structure of the solar upper atmosphere is
fundamental to constraining the coronal heating problem. Recent work has suggested that many
structures in the corona have relatively narrow distributions of temperatures
\cite[e.g.,][]{aschwanden2011,delzanna2015,warren2012,warren2008}. This result is difficult to
reconcile with the Parker nanoflare model of coronal heating \citep{parker1988}. All theoretical
calculations and numerical simulations indicate that magnetic reconnection occurs on spatial scales
that are much smaller than can be observed with current instrumentation. This implies that observed
temperature distributions should be broad \citep[e.g.,][]{cargill1994,klimchuk2001,cargill2004}.

Unfortunately, determining the temperature structure of the corona from remote sensing observations
is difficult. In principal, measurements of individual emission line intensities should yield this
information. The intensity, however, is a convolution of the temperature, density, and geometry
along the line of sight and such observations must be inverted to yield the differential emission
measure distribution (DEM). Since the inversion is ill posed it isn't clear what to make of the
solution. In their classic paper \citet{craig1976} write ``Consequently, in the derivation of
thermal structure from spectral data, observational errors are always magnified and often to such
an extent that the solution becomes meaningless.'' The inverse problem can be regularized so that
the solution is less sensitive to error but this introduces additional assumptions that may have no
physical basis \citep{judge1999}.

In this paper we introduce the application of sparse Bayesian inference to the differential
emission measure inversion problem. Here we are attempting to infer the temperature structure of
the atmosphere through observations of optically thin emission lines,
\begin{equation}
  {I_n} = \int \epsilon_n(n_e,T)\xi(T)\,dT,
\end{equation}
where $I_n$ is the observed intensity, $\epsilon_n(T)$ is the relevant plasma emissivity, which is
assumed to be known, and $\xi(T)=n_e^2\,ds/dT$ is the differential emission measure (DEM), which is
a function of the electron density and path length along the line of sight. As has been done in
many previous studies, we assume that the DEM can be represented by a sum of simple basis
functions. Motivated by the ``relevance vector machine,'' a Bayesian regression algorithm developed
by \citet{tipping2001}, we adopt a prior that encodes a preference for solutions that utilize a
minimum number of basis function. The important implication of this assumption is that the
complexity of the inferred temperature distribution is determined primarily by the observations and
their statistical significance and not by ad hoc assumptions about the solution.

To demonstrate the efficacy of this approach we have constructed a test library of 40 DEMs that
cover the range of what we expect to observe in the solar corona. For each distribution we estimate
the intensity and statistical uncertainty for a number of emission lines using the effective areas
of the Extreme Ultraviolet Imaging Spectrometer (EIS, \citealt{culhane2007}) on the \textit{Hinode}
mission. These intensities are used to attempt to recover the input DEM. We show that our method
outperforms another Bayesian DEM solver (MCMC by \citealt{kashyap1998}), which assumes an
uninformative prior for the weights.

This paper is structured in the following way. We first consider the application of our Bayesian
framework to a linear ``toy problem'' that closely follows example regression problems that are
often found in the literature \citep[e.g.,][]{tipping2004}. This allows us to make a gentle
introduction of the notation and to compare our approach with other established methods. We then
consider the application of our approach to the DEM inversion problem.

\section{The Linear Problem}

\begin{figure*}[t!]
 \centerline{%
   \includegraphics[angle=90,width=0.5\linewidth]{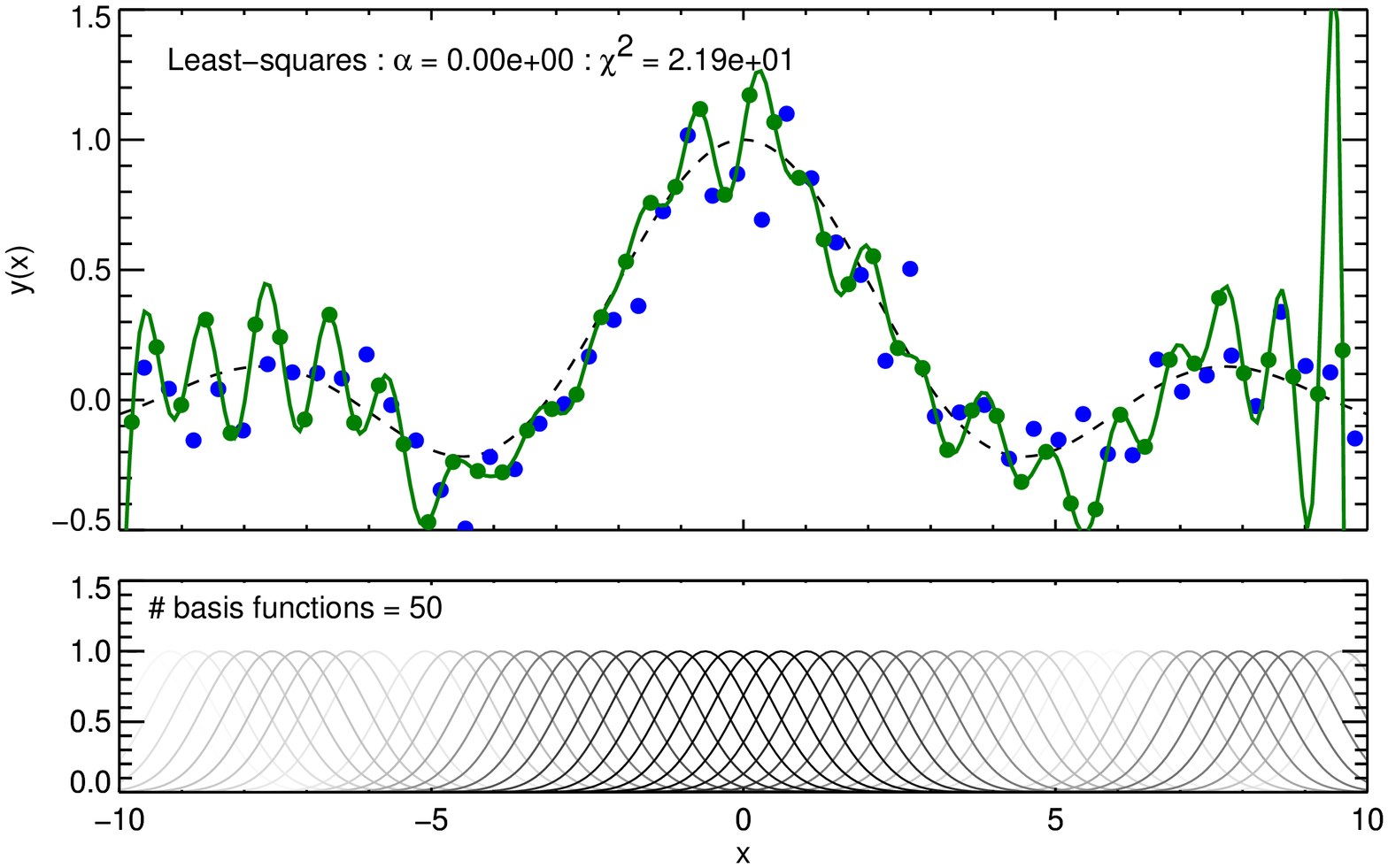}
   \includegraphics[angle=90,width=0.5\linewidth]{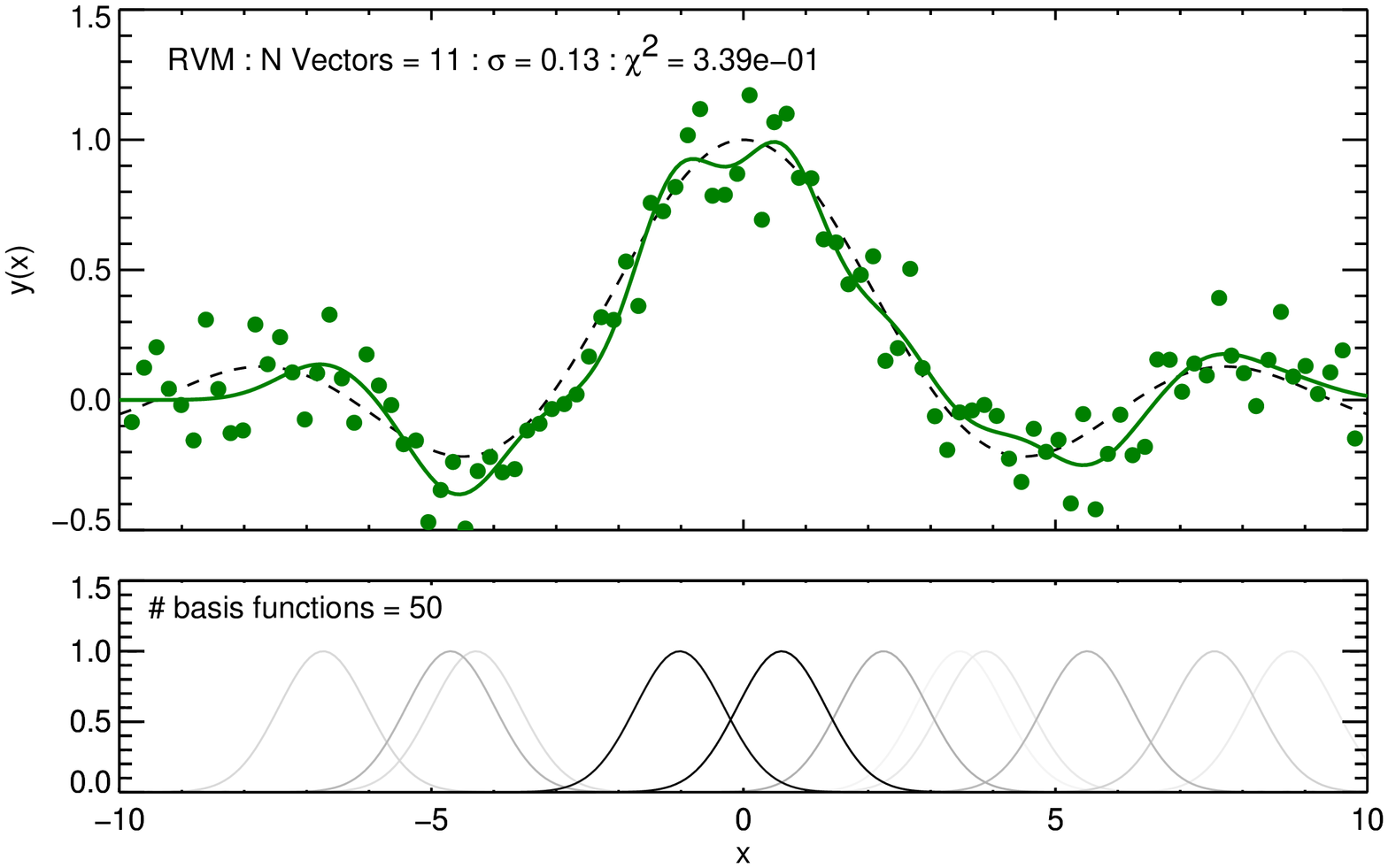}
   }
 \centerline{%
   \includegraphics[angle=90,width=0.5\linewidth]{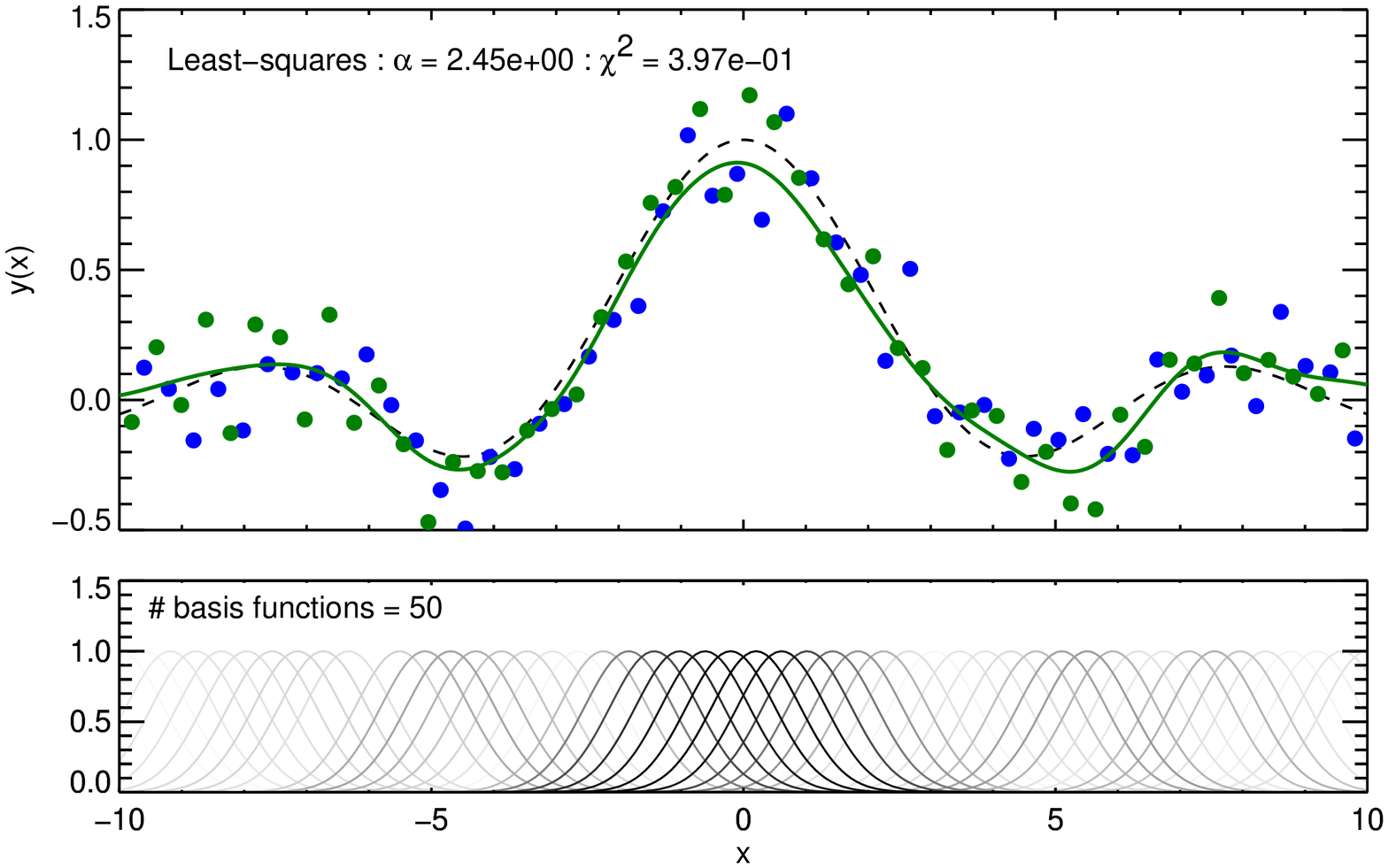}
   \includegraphics[angle=90,width=0.5\linewidth]{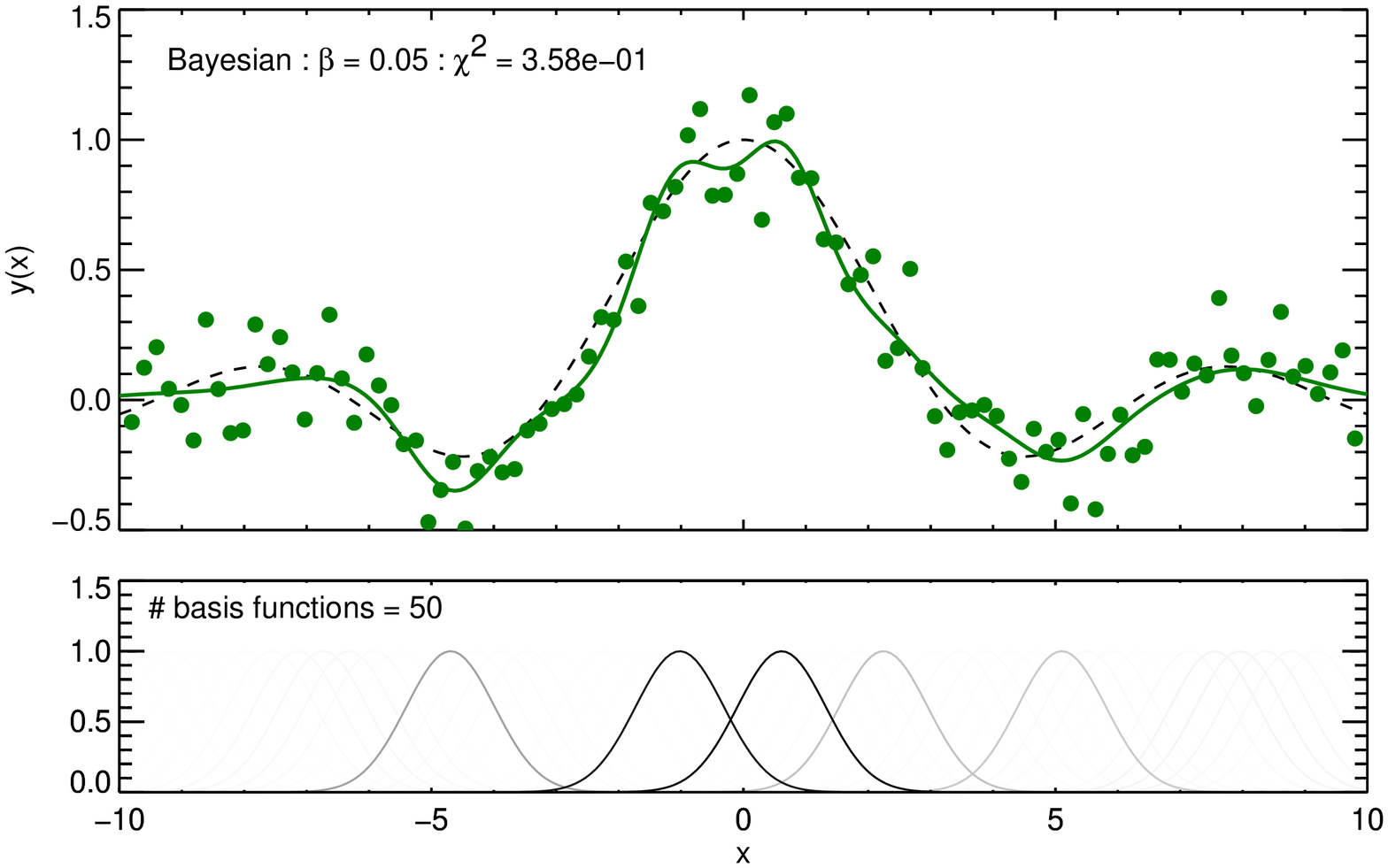}
   }
 \caption{Examples of regularizing a linear regression to noisy data generated from the
   $\mathrm{sinc}$ function. (\textit{top left}) A least-squares fit to half of the available data
   using 50 basis functions. This leads to overfitting of the training data and a poor
   representation of the test data. (\textit{lower left}) A penalized least-squares fit where the
   value of the regularization parameter $\alpha$ has been determined by cross-validation of the
   model and the test data. This leads to a solution that is smooth, but with many non-zero
   weights. (\textit{upper right}) An application of the \citet{tipping2001} relevance vector
   machine, a Bayesian formulation of the problem which iteratively removes basis functions to
   create a sparse representation. (\textit{lower right}) A Bayesian formulation of the problem
   where the prior, a Cauchy distribution, simply biases the weights to zero. The ``fat tails'' on
   the Cauchy distribution creates a preference for a set of weights where most are zero and only a
   few have finite values. Note that in each plot the basis functions are color coded according to
   their weight.} \label{fig:ToyProb}
\end{figure*}

In standard linear regression problems we wish to fit a set of $N$ observed data points ($x_n,
y_n$) to some smooth function with free parameters. We assume that the modeled data points
($\mathbf{\bar y}$) can be written as the linear superposition of a set of specified basis
functions
\begin{equation}
   {\bar y_n} = \sum_{m=1}^{M} \phi_m(x_n)w_m,
\end{equation}
and our task is to find the weights ($w_m$) that ``best fit'' the data. In this paper we will
assume that the basis functions are simple Gaussians
\begin{equation}
  \phi_m(x) = \exp\left[ -\frac{(x-x_m)^2}{r^2} \right]
\end{equation}
with a fixed width $r$. Anticipating the DEM problem we will not assume that the positions of the
basis functions ($x_m$) lie on the available data points but are evenly spaced on a fixed
domain.

For a given set of data the optimal weights can be found by minimizing the familiar expression
\begin{equation}
  \mathcal{L}  = \chi^2 = \frac{1}{2\sigma^2}\sum_{n=1}^{N} (y_n - {\bar y_n})^2,
  \label{eq:ls}
\end{equation}
for which the optimal weights can be found by gradient descent or Levenberg-Marquardt
\citep[e.g.,][]{markwardt2009}.  Also recall that we can cast this in matrix form
\begin{eqnarray}
  \mathbf{\bar y} & = & \mathbf{\Phi}\,\mathbf{w}, \\
  \mathcal{L} & = & \frac{1}{2\sigma^2}||\mathbf{y} -  \mathbf{\Phi}\,\mathbf{w}||^2,
\end{eqnarray}
where $\Phi_{mn} = \phi_m(x_n)$, which can be solved directly with
\begin{equation}
 \mathbf{w} = (\mathbf{\Phi}^\mathrm{T}\mathbf{\Phi})^{-1}\mathbf{\Phi}^\mathrm{T}\mathbf{y},
\end{equation}
or iteratively with gradient descent.

The challenge is to allow for many degrees of freedom so that we can fit a wide variety of
functions while not overfitting the data. To illustrate this, we generate 50 noisy data points from
the $\mathrm{sinc}(x)=\sin(x)/x$ function. The noise is drawn from a normal distribution with zero
mean and a standard deviation of $\sigma=0.15$. As is shown in Figure~\ref{fig:ToyProb}, a model
with with $M=50$ basis functions fits the data points perfectly, but is highly oscillatory. Our
intuition is that this model will not do a good job of predicting the values of new data
points. Indeed, if we generate another 50 noisy data points we see that $\chi^2$ is about an order
of magnitude larger for this new set.

The traditional approach to constraining such regression problems is penalized least squares, where
we minimize a function of the form
\begin{eqnarray}
  \mathcal{L} & = & \frac{1}{2\sigma^2}\sum_{n=1}^{N} (y_n - {\bar y_n})^2
  + \frac{\alpha}{2}\sum_{m=1}^{M}w_m^2 \label{eq:pls} \\
  \mathcal{L} & = & \frac{1}{2\sigma^2}||\mathbf{y} -  \mathbf{\Phi}\,\mathbf{w}||^2 +
  \frac{\alpha}{2}||\mathbf{w}||^2
\end{eqnarray}
which can be solved with gradient descent or directly with
\begin{equation}
  \mathbf{w} = (\mathbf{\Phi}^\mathrm{T}\mathbf{\Phi} + \alpha\mathbf{I})^{-1}
  \mathbf{\Phi}^\mathrm{T}\mathbf{y}.
\end{equation}

The parameter $\alpha$ balances the goodness of fit against the smoothness of the solution. It can
be found by using some of the data (the training data) to infer the weights and the remainder of
the data (the test data) to evaluate the goodness of fit. Since the computational complexity of the
problem is small, we can determine the value of $\alpha$ that best fits the test data through a
simple one-dimensional parameter search. Figure~\ref{fig:ToyProb} illustrates this approach to the
regression problem.

This method works well when we have many data points and can easily divide them into training and
test sets. Unfortunately, for the DEM problem we often have a limited number of observed
intensities and this ``leave some out'' cross-validation technique is not useful.

By reformulating the problem using a Bayesian framework we can achieve a similar result without
resorting to cross-validation. We write Bayes' theorem as the product of a likelihood and a
hierarchical prior
\begin{equation}
  p(\mathbf{w}|\mathbf{y}) = \frac{p(\mathbf{y}|\mathbf{w})p(\mathbf{w}|\beta)p(\beta)}
  {p(\mathbf{y})}.
  \label{eq:bayes}
\end{equation}
The likelihood is the usual expression assuming normally distributed errors on the observations
\begin{equation}
  p(\mathbf{y}|\mathbf{w}) = \prod_{n=1}^{N}\frac{1}{\sigma\sqrt{2\pi}}
  \exp\left[
    -\frac{(y_n - {\bar y_n})^2}{2\sigma^2} \right].
\end{equation}
For the prior on the weights we simply chose a Cauchy distribution,
\begin{equation}
  p(\mathbf{w}|\beta) = \prod_{m=1}^{M} \frac{\beta}{\pi}\frac{1}{\beta^2 + w_m^2},\quad \beta > 0.
  \label{eq:log_prior}
\end{equation}
We choose an uninformative prior for the hyperparameter $\beta$, which implicitly assumes that the
values for the weights will be only weakly dependent on the value of $\beta$ that we use. We
recognize that there are techniques for optimizing the hyperparameters and we will return to this
issue in the Summary and Discussion section.

If we are interested in the distributions of the weights implied by the data and our choice of the
posterior, we must generate samples from it. The standard approach to this is the
Metropolis-Hastings algorithm \citep{metropolis1953,hastings1970}. In the next section we will
describe the application of a powerful new parallel sampling technique
\citep{goodman2010,foreman2013,akeret2013} that we can apply to this problem. If we are only
interested in the set of weights that maximize the posterior, we take the negative log of the
posterior, discard all constant terms, and minimize
\begin{equation}
  \mathcal{L} = \frac{1}{2\sigma^2}\sum_{n=1}^{N} (y_n - {\bar y_n})^2
  + \sum_{m=1}^{M} \ln( \beta^2 + w_m^2), \label{eq:log_pos}
\end{equation}
which is similar to Equation~\ref{eq:pls}. This can be minimized using gradient descent.

Figure~\ref{fig:ToyProb} shows the solution assuming $\beta=0.05$ and using all 100 data points
simultaneously in the optimization. We see that this approach also avoids overfitting the data, but
by limiting the number of non-zero basis functions rather than by limiting the sum of the
weights. The Cauchy distribution encourages such sparse solutions because it has ``fat tails.'' To
see how this comes about consider a simple case where three basis functions could be used to
describe the data equally well. In the absence of a prior, the solution is likely to have weights
of comparable magnitude, e.g., $w_1, w_2, w_3 \sim \frac{1}{3}, \frac{1}{3}, \frac{1}{3}$, while we
would like the solution to be as simple as possible, say $w_1, w_2, w_3 \sim 1,0,0$. With the
assumption of the Cauchy prior, however, we have $p(1)p(0)p(0)\gg
p(\frac{1}{3})p(\frac{1}{3})p(\frac{1}{3})$ for $\beta\ll 1$ and we see that we can encode this
preference for models with a limited number of non-zero weights into the prior. Since the
temperature domain of the DEM is not unambiguously determined by the data, this preference for
sparse solutions is a useful property. We want emission measure to be inferred only when the
observations imply that it is statistically significant.

We note that this approach doesn't eliminate any degeneracy among the solutions. Several sets of
weights (e.g, 1,0,0 or 0,1,0 or 0,0,1 in our simple example) could be nearly equally likely. It
could also be that the numerical scheme used to explore the posterior has difficulty with such a
multi-modal landscape. We will discuss the problem of degeneracy in more detail in the next
section.

A more sophisticated Bayesian approach to sparsity has been formulated by \citet{tipping2001}, who
called his method the relevance vector machine (RVM). Here the prior on the weights is assumed to
be
\begin{equation}
  p(\mathbf{w}|\mathbf{\alpha}) = \prod_{m=1}^{M}\left[\frac{\alpha_m}{2\pi}\right]^{1/2}
              \exp\left[-\frac{\alpha_m w_m^2}{2}\right],
\end{equation}
where there is a hyperparameter $\alpha_m$ for each weight and the prior on each hyperparameter is
assumed to be a Gamma distribution. \citet{tipping2001} iteratively solves for the optimal set of
weights and hyperparameters by alternating between maximizing the likelihood with fixed $\alpha_m$
and maximizing the evidence (the denominator of Equation~\ref{eq:bayes}) with fixed $w_m$. As the iteration
proceeds, some of the $\alpha_m$'s become large and these basis functions are pruned from the
model. Upon convergence we are typically left with a model that has only a few remaining basis
functions.

Figure~\ref{fig:ToyProb} illustrates the application of this approach to our $\mathrm{sinc}$
problem. The RVM solution is similar to that obtained using the Cauchy prior, suggesting that this
distribution is consistent with sparsity. The Gamma function prior for $\alpha_m$ makes the
equivalent regularizing penalty proportional to $\sum \ln |w_m|$ (\citealt{tipping2001}, Equation
33), similar to what we obtain with the Cauchy prior, so perhaps this is not surprising.

\begin{figure*}[t!]
  \vspace{0.2in}
  \centerline{\includegraphics[width=0.80\linewidth]{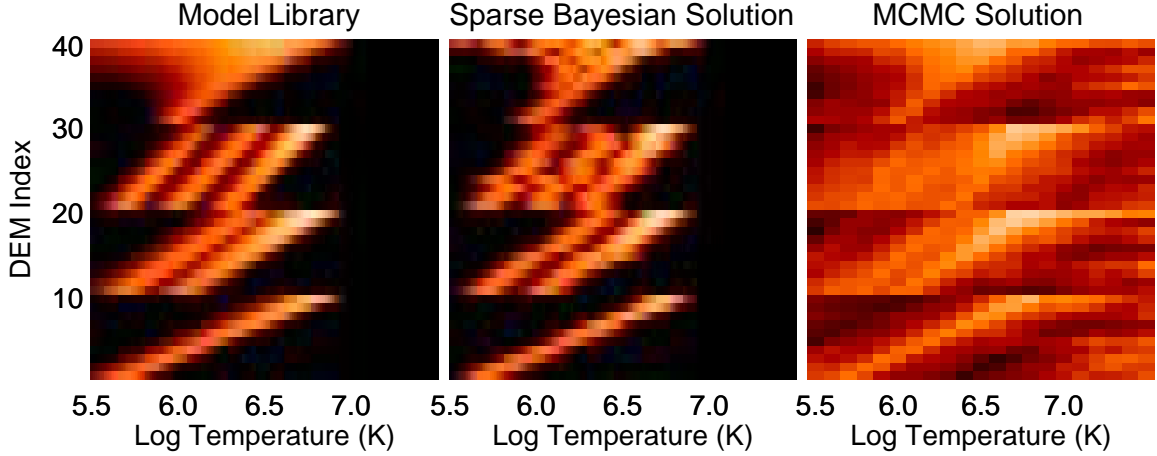}}
  \vspace{0.4in}
  \caption{The DEM library with 40 model DEMs along with the resulting solutions from the the
    Sparse Bayesian and MCMC methods. The online version of the paper contains machine readable
    tables for the DEM library and the reconstructed emission measure distributions, the final
    values of each weight, and the input and reconstructed intensities. }
  \label{fig:library_image}
\end{figure*}

For the scale of problem that we are interested in, the RVM is computationally
efficient. Unfortunately, the RVM allows for negative weights and cannot be used for the emission
measure problem since a negative emission measure has no physical meaning. The simple Bayesian
approach described by Equations~\ref{eq:bayes} and \ref{eq:log_pos}, in contrast, is much more
computationally expensive to implement, but it can easily accommodate the essential positive
definite constraint on the weights.

We close this section with a brief comment on mathematical rigor. It is clear that the further we
stray from simple least-squares solutions the less certain we are that our mathematical assumptions
will have their intended effect. We have certainly not proven that the Cauchy prior will always
lead to sparse solutions or that our assumptions about the hyper-parameter $\beta$ will not have
some perverse consequence in some circumstances. Similarly, \citet{tipping2001} has not proven that
the iterative approach to the RVM always converges to the optimal solution (for example, see
section of 2.2 of \citealt{candela2004} for cases where the RVM fails to perform optimally). Our
feeling is that considering a wide variety of test problems is the easiest way of dealing with this
lack of mathematical certainty.

\section{The DEM Problem}

We now turn to the actual problem of interest, inferring the temperature structure of the solar
atmosphere from observations of optically thin line emission. Specifically, we wish to invert the
equation
\begin{equation}
  {\bar I_n} = \int \epsilon_n(n_e,T)\xi(T)\,dT
\end{equation}
for the line-of-sight differential emission measure distribution $\xi(T)=n_e^2\,ds/dT$ given a set
of observed intensities ($I_n$), their associated statistical uncertainties ($\sigma_n$), and the
relevant plasma emissivites ($\epsilon_n(n_e,T))$. This problem differs from the linear problem in
that the function we wish to fit is not observed directly, but indirectly through the intensity.

As discussed in the previous section, the solution should have several properties: since the
density and path length are positive quantities, it must be positive definite; since in many
circumstances we will have only a limited number of observed intensities, it cannot rely on cross
validation methods for optimizing parameters; and it should be sparse, inferring emission measure
only at temperatures where there is a statistically significant reason to do so. As we will see,
the Bayesian approach that we applied to the linear problem satisfies all three of these
properties.

We assume that the product of the temperature and the DEM can be represented as the weighted sum of
known basis functions,
\begin{equation}
  T\xi(T) = \sum_{m=1}^M10^{w_m}\phi_m(T). \label{eq:dem}
\end{equation}
The exponential weight is chosen so that the DEM is positive definite. Note that since the range of
temperatures is large it is more efficient to integrate over intervals of constant log temperature
using $d\hspace{0.1em}\mathrm{ln}T = dT/T$. Putting the extra factor of $T$ on the right side of
Equation~\ref{eq:dem} allows both the weights and the basis functions to have uniform
magnitudes. For the basis functions we chose Gaussians in $\log T$,
\begin{equation}
  \phi_m(T) = 10^{20}\exp\left[-\frac{(\log T-\log T_m)^2}{2\sigma_T^2}\right].
\end{equation}
For a given temperature domain we assume that the basis functions are equally spaced and that the
width of each component is
\begin{equation}
\sigma_T = \frac{\log T_2-\log T_1}{2M},
\end{equation}
so that the width is automatically adjusted depending on the number of components.

\begin{deluxetable*}{rcrrrrr}
  \tablewidth{5in}
\tabletypesize{\scriptsize}
\tablecaption{Emission Lines of Interest and Example Calculation \tablenotemark{a}}
\tablehead{
  \multicolumn{1}{c}{Line} &
  \multicolumn{1}{c}{$\log T_{\max}$} &
  \multicolumn{1}{c}{Conversion} &
  \multicolumn{1}{c}{$I_{obs}$} &
  \multicolumn{1}{c}{$\sigma_{I}$} &
  \multicolumn{1}{c}{$I_{dem}$} &
  \multicolumn{1}{c}{$R$}
}
\startdata
    Mg V 276.579  &    5.45 &   2.06e-10  &     0.00  &   0.41  &     0.00 & 0.29 \\
   Mg VI 270.394  &    5.65 &   2.61e-10  &     0.00  &   0.32  &     0.00 & 1.79 \\
  Mg VII 280.737  &    5.80 &   1.38e-10  &     0.30  &   0.72  &     0.29 & 1.00 \\
  Si VII 275.368  &    5.80 &   2.24e-10  &     0.17  &   0.44  &     0.12 & 1.40 \\
   Fe IX 188.497  &    5.90 &   3.65e-10  &     1.72  &   0.74  &     1.37 & 1.26 \\
   Fe IX 197.862  &    5.90 &   6.44e-10  &     1.05  &   0.43  &     0.90 & 1.17 \\
    Fe X 184.536  &    6.05 &   1.54e-10  &    18.55  &   3.64  &    16.88 & 1.10 \\
    Si X 258.375  &    6.15 &   1.52e-10  &    62.36  &   5.65  &    62.28 & 1.00 \\
   Fe XI 180.401  &    6.15 &   3.98e-11  &   152.36  &  20.65  &   150.47 & 1.01 \\
   Fe XI 188.216  &    6.15 &   3.48e-10  &    76.93  &   4.84  &    75.87 & 1.01 \\
     S X 264.233  &    6.15 &   2.16e-10  &    11.25  &   2.02  &    11.37 & 0.99 \\
  Fe XII 195.119  &    6.20 &   7.11e-10  &   209.69  &   5.48  &   217.88 & 0.96 \\
  Fe XII 192.394  &    6.20 &   6.01e-10  &    67.57  &   3.41  &    70.20 & 0.96 \\
 Fe XIII 202.044  &    6.25 &   1.94e-10  &   115.64  &   7.68  &   120.15 & 0.96 \\
 Fe XIII 203.826  &    6.25 &   1.03e-10  &   174.43  &  12.90  &   182.51 & 0.96 \\
  Fe XIV 264.787  &    6.30 &   2.22e-10  &   243.24  &   9.07  &   245.37 & 0.99 \\
  Fe XIV 270.519  &    6.30 &   2.61e-10  &   124.92  &   5.94  &   125.95 & 0.99 \\
  Fe XIV 211.316  &    6.30 &   2.47e-11  &   456.62  &  41.82  &   459.89 & 0.99 \\
   Fe XV 284.160  &    6.35 &   9.54e-11  &  6114.98  &  66.95  &  6237.17 & 0.98 \\
  S XIII 256.686  &    6.40 &   1.35e-10  &   517.05  &  17.22  &   526.64 & 0.98 \\
  Fe XVI 262.984  &    6.45 &   2.02e-10  &  1028.86  &  19.62  &  1057.81 & 0.97 \\
  Ca XIV 193.874  &    6.55 &   6.73e-10  &   246.54  &   6.13  &   259.09 & 0.95 \\
  Ar XIV 194.396  &    6.55 &   6.93e-10  &    70.62  &   3.23  &    75.22 & 0.94 \\
   Ca XV 200.972  &    6.65 &   2.95e-10  &   210.21  &   8.40  &   213.79 & 0.98 \\
  Ca XVI 208.604  &    6.70 &   3.98e-11  &   120.19  &  17.08  &   115.32 & 1.04 \\
 Ca XVII 192.858  &    6.75 &   6.25e-10  &   157.95  &   5.10  &   138.12 & 1.14 \\
          AIA 94  &    6.85 &        ---  &    20.39  &   0.64  &    19.41 & 1.05
\enddata
\tablenotetext{a}{The conversion factor is the number of photons detected by EIS per unit radiance
  on the Sun and is used to estimate the statistical uncertainty for a modeled intensity. The
  observed intensities ($I_{obs}$) and statistical uncertainties ($\sigma_{I}$) are the inputs from
  the DEM library for DEM \#18. Computed intensities ($I_{dem}$) are from the Sparse Bayesian
  method.  The ratio R is $I_{obs}$/$I_{dem}$. The intensities are in units of
  erg cm$^{-2}$ s$^{-1}$ sr$^{-1}$.}
\label{table:line_list}
\end{deluxetable*}

With these assumptions it is possible to integrate the emissivity over each basis function,
\begin{align}
  {\bar I_n} & =  \int \epsilon_n(T)\,T\xi(T)\,d\hspace{0.1em}
         \mathrm{ln}T, \label{eq:ints} \\
  {\bar I_n} & =  \int \epsilon_n(T)\sum_{m=1}^M10^{w_m}\phi_m(T)\,d\hspace{0.1em}
         \mathrm{ln}T, \\
  {\bar I_n} & =  \sum_{m=1}^M10^{w_m} \int \epsilon_n(T)\phi_m(T)\,d\hspace{0.1em}
         \mathrm{ln}T, \label{eq:design} \\
  {\bar I_n} & =  \sum_{m=1}^M 10^{w_m} \phi_{mn},
\end{align}
and the calculation of each modeled intensity is reduced to a simple sum. The only differences
between this problem and the problem discussed in the previous section are that the intensities are
non-linear functions of the weights and, since spectroscopic instruments rarely have uniform
sensitivity, each intensity has a uncertainty ($\sigma_n$).

The line-of-sight emission measures observed on the Sun are typically large,
$\sim10^{27}$\,cm$^{-5}$ in an active region, for example \citep{warren2012}.  To bias the weights
to the domain of interest we introduce a scaling factor into the basis functions so that a weight
of zero corresponds to $10^{20}$ instead of 1. Further we modify the prior on the weights
(Equation~\ref{eq:log_prior}) by mapping $w_m^2$ to $(w_m-20)^2$.


Given a set of observed intensities and the corresponding atomic data, we can find the optimal
values for the weights either by sampling the posterior (Equation~\ref{eq:bayes}) or minimizing the
negative log-posterior (Equation \ref{eq:log_pos}). Unfortunately, finding global extrema in a high
dimensional space is a formidable problem. Our approach is to begin by sampling the posterior using
the method introduced by \citet{goodman2010} and implemented by \citet{foreman2013} and
\citet{akeret2013}, among others. This method differs from the traditional single Monte Carlo
Markov chain random walk method for sampling the posterior by utilizing an ensemble of walkers,
potentially thousands of them, that can be updated in parallel. Further, instead of using a
proposal distribution with parameters that must be adjusted to achieve the desired acceptance
ratio, updates to the chain are made using the current positions of pairs of walkers
\begin{equation}
  \mathbf{w}_{t+1}^j = \mathbf{w}_{t}^i + z(\mathbf{w}_{t}^j - \mathbf{w}_{t}^i),
\end{equation}
where $z$ is a random number from the distribution $g(z) = 1/\sqrt{z}$ on the domain $[1/a, a]$. In
principle, $a$ is an adjustable parameter but the choice $a=2$ appears to work well in all
applications \citep{foreman2013}. Note that the pairs of walkers are shuffled on each iteration.

We implemented this algorithm in a simple C code that was parallelized using openMP. We validated
it by sampling from known posteriors, applying it to simple parameter estimation problems, such as
fitting spectral profiles with a Gaussian, and using it for the linear regression problem in the
previous section.

\begin{figure*}[t!]
  \centerline{\includegraphics[angle=90,width=\linewidth]{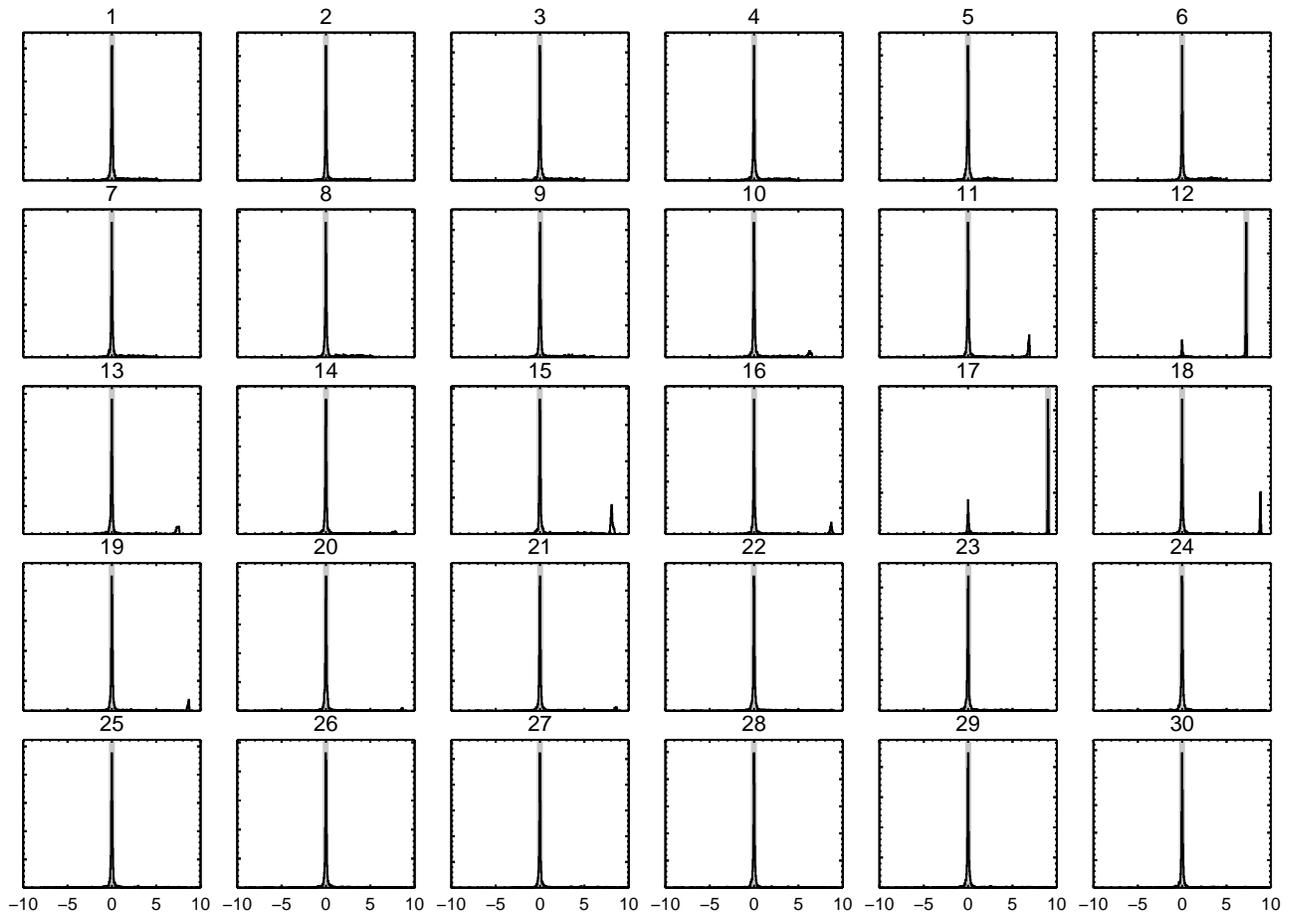}}
  \caption{The posterior distributions of weights for each of the walkers from the sampling for
    library DEM \#18. While these distributions have multiple peaks, using the mode of each
    distribution generally recovers the input DEM (see Figure~\ref{fig:library_evo}). The mode is
    indicated as a gray line behind each posterior distribution.}
  \label{fig:library_dist}
\end{figure*}

To test our approach we need to create some DEMs and select a line list for our hypothetical
observations. In Figure~\ref{fig:library_image} we show the ``library'' of 40 test DEMs that we
have devised. These functions are superpositions of Gaussians that are either linear or logarithmic
in temperature and range from narrow to very broad, representing the range of curves that have been
found in previous studies \citep[e.g.,][]{feldman1998,warren2013}. Note that we have deliberately
mismatched the widths of the DEMs in the library and the widths of the basis functions.

To complement the DEM library we select a series of emission lines that are observed with EIS and
that are typically used in temperature studies \citep[e.g.,][]{warren2011}. These lines are listed
in Table~\ref{table:line_list}, where we also list the peak temperature of formation for the ion
and the conversion from the radiance observed at the sun to photons detected by the instrument,
which is needed to compute the statistical uncertainty for each intensity. This conversion factor
is simply the product of the pre-flight effective area \citep{lang2006}, the solid angle subtended
by a pixel, and the exposure time.  Here we assume the use of the 1\arcsec\ slit and a 100\,s
exposure time.

To augment the EIS line list we use a calculated intensity for the 94\,\AA\ channel on the the
Atmospheric Imaging Assembly (AIA, \citealt{lemen2012}). This channel can observe \ion{Fe}{18}
93.932\,\AA, which peaks at about $\log T = 6.85$. We assume that we are using 94\,\AA\ images that
have been processed to remove lower temperature emission (see the appendix of
\citealt{warren2012}). Errors on the intensity are computed using the standard analysis routine
\verb+aia_bp_estimate_error+.

For all of our emissivity calculations we use the CHIANTI atomic database version 8.0.1
\citep{delzanna2015b,dere1997} assuming coronal abundances \citep{feldman1992}, a constant pressure
of $P_e = 2n_eT_e = 10^{16}$\,cm$^{-3}$~K, and the default ionization fractions. For each emission
line the emissivities of all lines with 0.5\,\AA\ are summed to create a composite response.

We are now in a position to calculate the DEMs. To compute the intensities we convolve each
distribution in the library with the emissivities for all of the EIS lines and the AIA
94\,\AA\ channel. We also compute the statistical uncertainty for each of the resulting
intensities. For the prior we use $\beta=0.05$. For the sampling we initialize 1,000 walkers with
randomly selected weights on the interval $[0.1, 9.0]$ and begin iterating. After a burn-in of
$10^7$ accepted samples, the next $10^4$ proposed steps that are accepted are recorded to form the
final sampling of the posterior. After each iteration we also recorded the highest value of the
posterior and saved the corresponding set of weights. This procedure takes about 45\,min per DEM on
a standard 4-core 4\,GHz Intel i7 processor with all four cores utilized. We will discuss how this
algorithm could be made more efficient in the next section.

In Figure~\ref{fig:library_dist} we show the distributions of the weights obtained for one of the
DEM calculations (\#18). The corresponding intensities are shown in
Table~\ref{table:line_list}. Unfortunately, these posterior distributions are not simple, single
peaked functions, but have multiple peaks. As discussed in the previous section, multiple peaks are
an inevitable consequence of having many basis functions. Recall our simple example of having three
degenerate basis functions where each could represent the data equally well. The posterior in this
simple case would consist of approximately equal parts (1,0,0), (0,1,0), and (0,0,1) and the
distribution for each weight would have a strong peak at 0 and a secondary peak at 1.  For the DEM
problem we have not specified so many basis functions that we expect to see such extreme
degeneracy. Still, multiple peaks are clearly present in the posterior distributions for many
weights.

\begin{figure}[t!]
  \centerline{\includegraphics[width=0.84\linewidth]{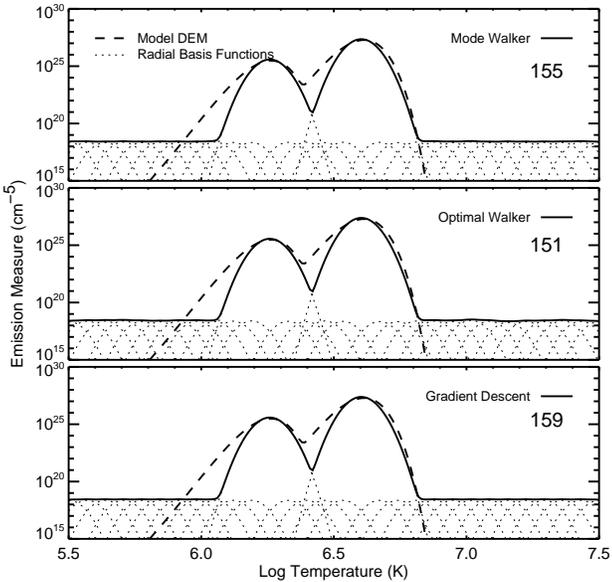}}
  \vspace{0.2in}
  \caption{Computed solutions for library DEM \#18 using the mode of the posterior weight
    distributions, the walker with the highest observed value of the posterior for all accepted
    samples, and the solution using this walker as an initial condition for gradient descent. In
    all cases the largest value of the posterior is obtained by gradient descent. The value of the
    un-normalized log-posterior in each case is indicated in the figure (larger is better). Note
    that in this and all subsequent plots we show the quantity $T\xi(T)d\ln T$.}
  \label{fig:library_evo}
\end{figure}

\begin{figure}[t!]
  \centerline{\includegraphics[width=0.84\linewidth]{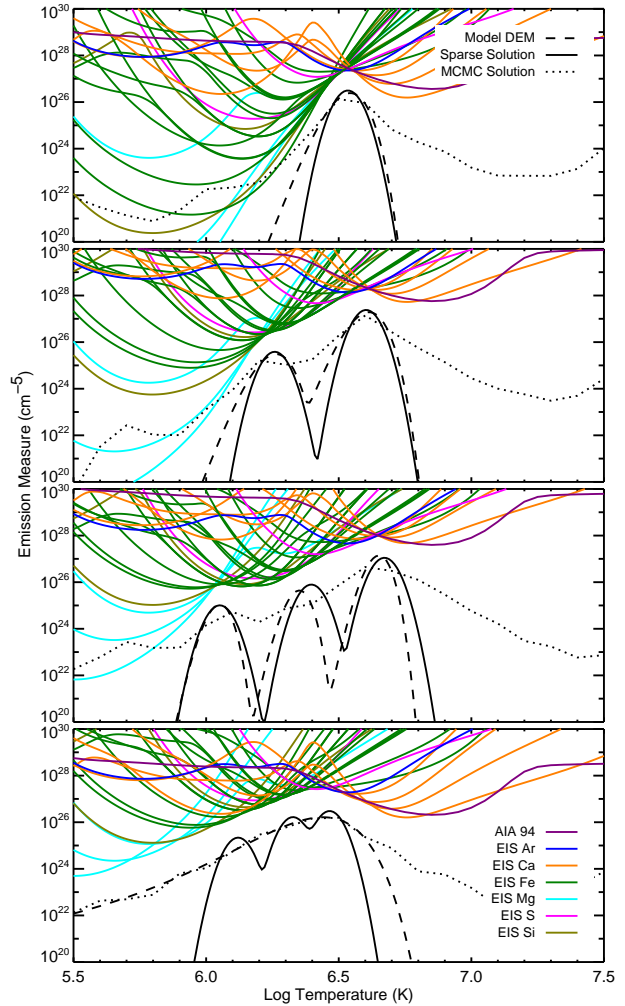}}
  \vspace{0.2in}
  \caption{Computed results for library DEM \#8, \#18, \#28, and \#38 showing the model
    input, Sparse Bayesian solution, and MCMC solution. The colored curves are the emission measure
    loci curves ($I_n/\epsilon_\lambda(T)$) and form an envelope on the emission measure
    distribution.}
  \label{fig:library_dems}
\end{figure}

\begin{figure}[t!]
  \centerline{\includegraphics[width=\linewidth]{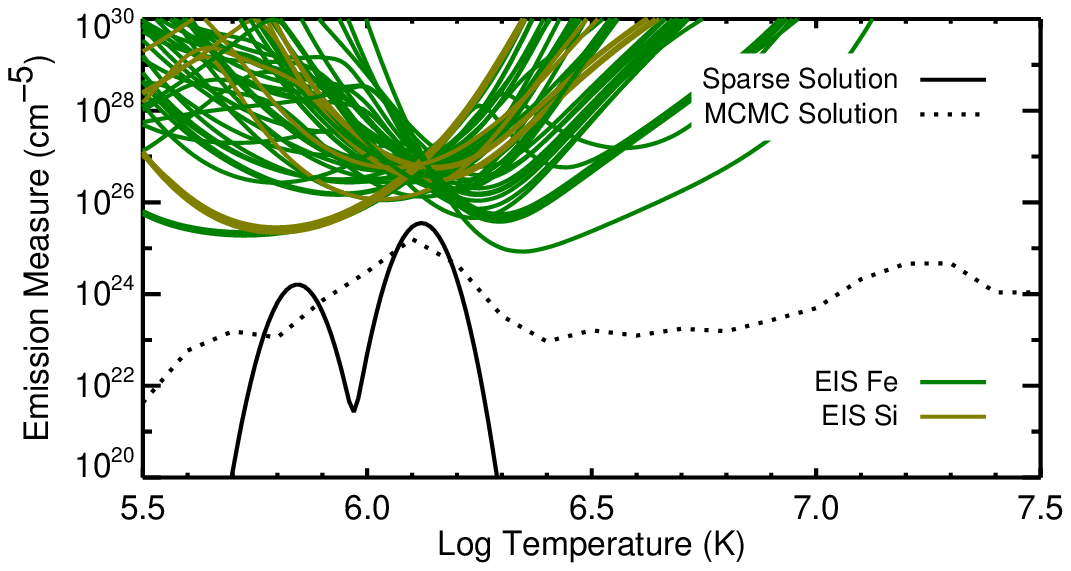}}
  \vspace{-0.15in}
  \centerline{\includegraphics[width=\linewidth]{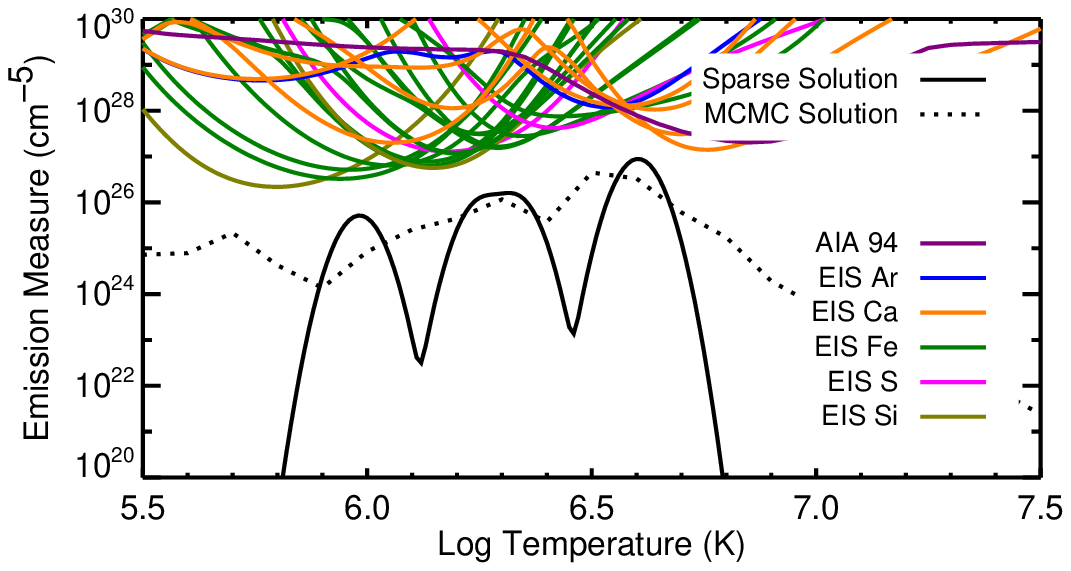}}
  \caption{Sparse Bayesian DEMs computed for two sets of observed EIS intensities.
    \textit{(top panel)} The off-limb quiet sun intensities from \citet{warren2014}.
    \textit{(bottom panel)} The active region intensities from \citet{warren2012}.}
  \label{fig:observed}
\end{figure}

So which weights do we use to compute the final DEM? The median and mean of the weight
distributions are potentially problematic. The mode of the distribution, however, is generally well
behaved. Using these weights to compute the DEM both reproduces the observed intensities and the
input DEM very closely. For each run we also used the walker with the highest value of the
posterior as initial conditions for a gradient descent calculation. As is illustrated in
Figure~\ref{fig:library_evo}, the DEMs recovered using gradient descent generally matched those
recovered using the mode of the posterior weight distribution. In all cases the weights obtained
after gradient descent actually produced a higher value for the posterior than the mode, suggesting
that using the mode is not optimal. For this reason we use the weights obtained from gradient
descent to represent the recovered DEM.

We have performed this calculation for all of the DEMs in the library. Some example DEMs are shown
in Figure~\ref{fig:library_dems}. The summary of all of the calculations is presented in
Figure~\ref{fig:library_image}. In almost all cases the input intensities are reproduced to within
a few percent and the general properties of the input DEM are also recovered and no emission
measure is inferred far away from the peaks in the input distributions.

We note that the excellent agreement between the recovered and actual DEMs is only achieved after a
very long burn-in period (recall that we accepted $10^7$ samples before collecting the samples for
the posterior). For a burn-in of $10^5$ samples the mode of the posterior distribution resulted in
a poor approximation of the input DEM with finite weights for most of basis functions. For a
burn-in of $10^6$ samples the recovered DEMs were generally close to the input DEMs but there were
a number of spurious peaks. Many of these peaks remained even after performing gradient descent. 

The sampling of the posterior is computationally expensive and one might wonder if it would be
easier to simply select some random initial conditions and then do gradient descent. Our
experiments with this were not encouraging. In almost all cases the calculation converged to some
local maximum far from the solution that we obtained through sampling. We note that we did not
study this extensively. Our intuition, however, is that the sampling is an invaluable aid in
exploring a complex posterior in a high dimensional space and is worth the computational expense.

To provide a point of comparison for our calculations we have used another Bayesian solver, the
``MCMC'' routine of \citealt{kashyap1998}, on each set of computed intensities in the DEM
library. Here we use 20 temperature bins in $\log T$ and run 100 simulations on each DEM.
The resulting DEMs, which are summarized in Figure~\ref{fig:library_image}, also recovers
the main features of each input distribution. We also see, however, that MCMC infers emission
measure at temperatures far away from the main peaks in the input distributions.  This is not
surprising given the assumption of an uninformative prior over the weights.

Finally, we have also applied this technique to two sets of observed EIS intensities. These results
are summarized in Figure~\ref{fig:observed}. The sparse technique that we introduced here recovers
the observed intensities as well as MCMC. Since the DEM library was constructed with these types of
DEMs in mind -- narrow DEMs for off-limb quiet Sun and somewhat broader DEMs for active regions --
it is not surprising that the algorithm can recover them.

\section{Summary and Discussion}

We have explored the application of sparse Bayesian inference to the problem of determining the
temperature structure of the solar corona. We have found that by adopting a Cauchy distribution as
a prior for the weights we can encode a preference for a sparse representation of the temperature
distribution. Thus we can use a rich set of basis functions and recover a wide variety of DEMs,
while limiting the amount of spurious emission measure at other temperatures. We have also shown
that a complex, high-dimensional posterior can be explored in detail using the parallel sampling
technique of \citet{goodman2010}. The results from this sparse Bayesian approach compare favorably
to the results obtained from MCMC \citep{kashyap1998}. As one would expect, the absence of any
constraints on the weights in MCMC leads to considerable emission measure being inferred at
temperatures away from the peaks in the input DEMs.

There are two aspects of our algorithm that could be improved. First, we have not considered the
optimization of the hyperparameters. The magnitude of the weight penalty ($\beta$ in
Equation~\ref{eq:log_prior}), the number of basis functions, and the widths of the basis functions,
are left fixed. As mentioned in the discussion of the RVM in the previous section, one method for
optimizing the hyperparameters is maximizing the evidence (see Chapter 18 of \citealt{barber2012}
for more details). Unfortunately, our non-linear representation of the DEM makes the integration of
the evidence analytically intractable. A linear representation of the DEM would make many aspects
of the problem simpler, but this would require an algorithm for optimizing the weights that
incorporates the positive definite constraint. We conjecture that such an approach would be orders
of magnitude faster than our current approach. 

We have limited our comparisons to MCMC. There have been many papers written on DEM inversions, but
a complete summary of techniques is well beyond the scope of this work. Two recent efforts,
however, directly address some of the issues that we have discussed in this paper. \citet{hannah2012}
implement a penalized least squares algorithm (Equation~\ref{eq:pls}) where the hyperparmeter
$\alpha$ is adjusted using the statistical uncertainty in observations and the non-negativity
constraint. As illustrated with our linear problem, this produces solutions that are smooth, but
not necessarily sparse, and thus sensitive to the domain specified for the inversion. One typically
chooses a temperature range that extends slightly beyond the temperature of formation of the
coolest and hottest line that is observed. This is not a particularly controversial assumption and
so this is not a major shortcoming. \citet{cheung2015} have implemented an algorithm with a weight
penalty based on the L1 norm \citep[e.g.,][]{tibshirani1996,candes2006}. \citet{cheung2015} do not
address the optimization of the hyperparmeter on the weight penalty. They also use a L1 norm for
the likelihood and it is not clear if this is consistent with the uncertainties of the
observations. Both \citet{hannah2012} and \citet{cheung2015} have tested their algorithms on a wide
variety of input DEMs and they perform well. These algorithms are significantly faster than our
sparse Bayesian method.

Finally, we stress that our goals for the DEM are relatively modest. In general, we simply wish to
understand if structures in the solar atmosphere have narrow or broad temperature distributions. We
also stress that no amount of mathematical machinery can replace the need for well-calibrated
observations that contain emission lines that cover a wide range of temperatures. Highly accurate
atomic data is also a critical component of any effort to understand the temperature structure of
the solar atmosphere \citep[e.g.,][]{guennou2013}.


\acknowledgments This work was sponsored by the Chief of Naval Research and NASA's Hinode
project. Hinode is a Japanese mission developed and launched by ISAS/JAXA, with NAOJ as domestic
partner and NASA and STFC (UK) as international partners.



\begin{thebibliography}{}
\expandafter\ifx\csname natexlab\endcsname\relax\def\natexlab#1{#1}\fi

\bibitem[{{Akeret} {et~al.}(2013){Akeret}, {Seehars}, {Amara}, {Refregier}, \&
  {Csillaghy}}]{akeret2013}
{Akeret}, J., {Seehars}, S., {Amara}, A., {Refregier}, A., \& {Csillaghy}, A.
  2013, Astronomy and Computing, 2, 27

\bibitem[{{Aschwanden} \& {Boerner}(2011)}]{aschwanden2011}
{Aschwanden}, M.~J., \& {Boerner}, P. 2011, \apj, 732, 81

\bibitem[{Barber(2012)}]{barber2012}
Barber, D. 2012, Bayesian Reasoning and Machine Learning (New York, NY, USA:
  Cambridge University Press)

\bibitem[{Candela(2004)}]{candela2004}
Candela, J.~Q. 2004, PhD thesis, Informatics and Mathematical Modelling,
  Technical University of Denmark, {DTU}, Richard Petersens Plads, Building
  321, {DK-}2800 Kgs. Lyngby, supervised by Professor Lars Kai Hansen

\bibitem[{{Cand{\`e}s} {et~al.}(2006){Cand{\`e}s}, {Romberg}, \&
  {Tao}}]{candes2006}
{Cand{\`e}s}, E., {Romberg}, J., \& {Tao}, T. 2006, IEEE Transactions on
  Information Theory, 52, 489

\bibitem[{{Cargill}(1994)}]{cargill1994}
{Cargill}, P.~J. 1994, \apj, 422, 381

\bibitem[{{Cargill} \& {Klimchuk}(2004)}]{cargill2004}
{Cargill}, P.~J., \& {Klimchuk}, J.~A. 2004, \apj, 605, 911

\bibitem[{{Cheung} {et~al.}(2015){Cheung}, {Boerner}, {Schrijver}, {Testa},
  {Chen}, {Peter}, \& {Malanushenko}}]{cheung2015}
{Cheung}, M.~C.~M., {Boerner}, P., {Schrijver}, C.~J., {et~al.} 2015, \apj,
  807, 143

\bibitem[{{Craig} \& {Brown}(1976)}]{craig1976}
{Craig}, I.~J.~D., \& {Brown}, J.~C. 1976, \aap, 49, 239

\bibitem[{{Culhane} {et~al.}(2007){Culhane}, {Harra}, {James}, {Al-Janabi},
  {Bradley}, {Chaudry}, {Rees}, {Tandy}, {Thomas}, {Whillock}, {Winter},
  {Doschek}, {Korendyke}, {Brown}, {Myers}, {Mariska}, {Seely}, {Lang}, {Kent},
  {Shaughnessy}, {Young}, {Simnett}, {Castelli}, {Mahmoud}, {Mapson-Menard},
  {Probyn}, {Thomas}, {Davila}, {Dere}, {Windt}, {Shea}, {Hagood}, {Moye},
  {Hara}, {Watanabe}, {Matsuzaki}, {Kosugi}, {Hansteen}, \&
  {Wikstol}}]{culhane2007}
{Culhane}, J.~L., {Harra}, L.~K., {James}, A.~M., {et~al.} 2007, \solphys, 243,
  19

\bibitem[{{Del Zanna} {et~al.}(2015{\natexlab{a}}){Del Zanna}, {Dere}, {Young},
  {Landi}, \& {Mason}}]{delzanna2015b}
{Del Zanna}, G., {Dere}, K.~P., {Young}, P.~R., {Landi}, E., \& {Mason}, H.~E.
  2015{\natexlab{a}}, \aap, 582, A56

\bibitem[{{Del Zanna} {et~al.}(2015{\natexlab{b}}){Del Zanna}, {Tripathi},
  {Mason}, {Subramanian}, \& {O'Dwyer}}]{delzanna2015}
{Del Zanna}, G., {Tripathi}, D., {Mason}, H., {Subramanian}, S., \& {O'Dwyer},
  B. 2015{\natexlab{b}}, \aap, 573, A104

\bibitem[{{Dere} {et~al.}(1997){Dere}, {Landi}, {Mason}, {Monsignori Fossi}, \&
  {Young}}]{dere1997}
{Dere}, K.~P., {Landi}, E., {Mason}, H.~E., {Monsignori Fossi}, B.~C., \&
  {Young}, P.~R. 1997, \aaps, 125, 149

\bibitem[{{Feldman} {et~al.}(1992){Feldman}, {Mandelbaum}, {Seely}, {Doschek},
  \& {Gursky}}]{feldman1992}
{Feldman}, U., {Mandelbaum}, P., {Seely}, J.~F., {Doschek}, G.~A., \& {Gursky},
  H. 1992, \apjs, 81, 387

\bibitem[{{Feldman} {et~al.}(1998){Feldman}, {Sch{\"u}hle}, {Widing}, \&
  {Laming}}]{feldman1998}
{Feldman}, U., {Sch{\"u}hle}, U., {Widing}, K.~G., \& {Laming}, J.~M. 1998,
  \apj, 505, 999

\bibitem[{{Foreman-Mackey} {et~al.}(2013){Foreman-Mackey}, {Hogg}, {Lang}, \&
  {Goodman}}]{foreman2013}
{Foreman-Mackey}, D., {Hogg}, D.~W., {Lang}, D., \& {Goodman}, J. 2013, \pasp,
  125, 306

\bibitem[{{Goodman} \& {Weare}(2010)}]{goodman2010}
{Goodman}, J., \& {Weare}, J. 2010, Commun. Appl. Math. Comput. Sci., 5, 65

\bibitem[{{Guennou} {et~al.}(2013){Guennou}, {Auch{\`e}re}, {Klimchuk},
  {Bocchialini}, \& {Parenti}}]{guennou2013}
{Guennou}, C., {Auch{\`e}re}, F., {Klimchuk}, J.~A., {Bocchialini}, K., \&
  {Parenti}, S. 2013, \apj, 774, 31

\bibitem[{{Hannah} \& {Kontar}(2012)}]{hannah2012}
{Hannah}, I.~G., \& {Kontar}, E.~P. 2012, \aap, 539, A146

\bibitem[{{Hastings}(1970)}]{hastings1970}
{Hastings}, W.~K. 1970, Biometrika, 57, 97

\bibitem[{{Judge} \& {McIntosh}(1999)}]{judge1999}
{Judge}, P.~G., \& {McIntosh}, S.~W. 1999, \solphys, 190, 331

\bibitem[{{Kashyap} \& {Drake}(1998)}]{kashyap1998}
{Kashyap}, V., \& {Drake}, J.~J. 1998, \apj, 503, 450

\bibitem[{{Klimchuk} \& {Cargill}(2001)}]{klimchuk2001}
{Klimchuk}, J.~A., \& {Cargill}, P.~J. 2001, \apj, 553, 440

\bibitem[{{Lang} {et~al.}(2006){Lang}, {Kent}, {Paustian}, {Brown}, {Keyser},
  {Anderson}, {Case}, {Chaudry}, {James}, {Korendyke}, {Pike}, {Probyn},
  {Rippington}, {Seely}, {Tandy}, \& {Whillock}}]{lang2006}
{Lang}, J., {Kent}, B.~J., {Paustian}, W., {et~al.} 2006, \ao, 45, 8689

\bibitem[{{Lemen} {et~al.}(2012){Lemen}, {Title}, {Akin}, {Boerner}, {Chou},
  {Drake}, {Duncan}, {Edwards}, {Friedlaender}, {Heyman}, {Hurlburt}, {Katz},
  {Kushner}, {Levay}, {Lindgren}, {Mathur}, {McFeaters}, {Mitchell}, {Rehse},
  {Schrijver}, {Springer}, {Stern}, {Tarbell}, {Wuelser}, {Wolfson}, {Yanari},
  {Bookbinder}, {Cheimets}, {Caldwell}, {Deluca}, {Gates}, {Golub}, {Park},
  {Podgorski}, {Bush}, {Scherrer}, {Gummin}, {Smith}, {Auker}, {Jerram},
  {Pool}, {Soufli}, {Windt}, {Beardsley}, {Clapp}, {Lang}, \&
  {Waltham}}]{lemen2012}
{Lemen}, J.~R., {Title}, A.~M., {Akin}, D.~J., {et~al.} 2012, \solphys, 275, 17

\bibitem[{{Markwardt}(2009)}]{markwardt2009}
{Markwardt}, C.~B. 2009, in Astronomical Society of the Pacific Conference
  Series, Vol. 411, Astronomical Data Analysis Software and Systems XVIII, ed.
  D.~A. {Bohlender}, D.~{Durand}, \& P.~{Dowler}, 251

\bibitem[{{Metropolis} {et~al.}(1953){Metropolis}, {Rosenbluth}, {Rosenbluth},
  {Teller}, \& {Teller}}]{metropolis1953}
{Metropolis}, N., {Rosenbluth}, A.~W., {Rosenbluth}, M.~N., {Teller}, A.~H., \&
  {Teller}, E. 1953, J. Chem. Phys., 21, 1087

\bibitem[{{Parker}(1988)}]{parker1988}
{Parker}, E.~N. 1988, \apj, 330, 474

\bibitem[{Tibshirani(1996)}]{tibshirani1996}
Tibshirani, R. 1996, J. Royal Stat. Soc. B, 58, 267

\bibitem[{Tipping(2001)}]{tipping2001}
Tipping, M.~E. 2001, J. Mach. Learn. Res., 1, 211

\bibitem[{Tipping(2004)}]{tipping2004}
---. 2004, Bayesian Inference: An Introduction to Principles and Practice in
  Machine Learning, ed. O.~Bousquet, U.~von Luxburg, \& G.~R{\"a}tsch (Berlin,
  Heidelberg: Springer Berlin Heidelberg), 41--62

\bibitem[{{Warren} {et~al.}(2011){Warren}, {Brooks}, \&
  {Winebarger}}]{warren2011}
{Warren}, H.~P., {Brooks}, D.~H., \& {Winebarger}, A.~R. 2011, \apj, 734, 90

\bibitem[{{Warren} {et~al.}(2013){Warren}, {Mariska}, \&
  {Doschek}}]{warren2013}
{Warren}, H.~P., {Mariska}, J.~T., \& {Doschek}, G.~A. 2013, \apj, 770, 116

\bibitem[{{Warren} {et~al.}(2008){Warren}, {Ugarte-Urra}, {Doschek}, {Brooks},
  \& {Williams}}]{warren2008}
{Warren}, H.~P., {Ugarte-Urra}, I., {Doschek}, G.~A., {Brooks}, D.~H., \&
  {Williams}, D.~R. 2008, \apjl, 686, L131

\bibitem[{{Warren} {et~al.}(2014){Warren}, {Ugarte-Urra}, \&
  {Landi}}]{warren2014}
{Warren}, H.~P., {Ugarte-Urra}, I., \& {Landi}, E. 2014, \apjs, 213, 11

\bibitem[{{Warren} {et~al.}(2012){Warren}, {Winebarger}, \&
  {Brooks}}]{warren2012}
{Warren}, H.~P., {Winebarger}, A.~R., \& {Brooks}, D.~H. 2012, \apj, 759, 141

\end{thebibliography}
\end{document}